\documentclass[twocolumn,showpacs,preprintnumbers,amsmath,amssymb]{revtex4}
\usepackage{epsfig}

\newtheorem{theorem}{Theorem}

\begin{document}


\title{Local Permutations of Products of Bell States\\
       and Entanglement Distillation}

\author{Jeroen Dehaene}
\email{Jeroen.Dehaene@esat.kuleuven.ac.be}
\author{Maarten Van den Nest}
\affiliation{Katholieke Universiteit Leuven, ESAT-SCD, Belgium}
\author{Frank Verstraete}
\altaffiliation[Also at ]{K.U.Leuven, ESAT-SCD}
\affiliation{Universiteit Gent,
             Department of Mathematical Physics and Astronomy, Belgium}
\author{Bart De Moor}
\affiliation{Katholieke Universiteit Leuven, ESAT-SCD, Belgium}
\date{\today}

\begin{abstract}
We present new algorithms for mixed-state multi-copy entanglement
distillation for pairs of qubits. Our algorithms perform significantly
better than the best known algorithms. Better algorithms can be
derived that are tuned for specific initial states.
The new algorithms are based on a characterization of the group of all
locally realizable permutations of the $4^n$ possible tensor products
of $n$ Bell states.
\end{abstract}


\maketitle

\section{Introduction}

We study mixed-state multi-copy entanglement distillation protocols
for pairs of qubits. We start from $n$ identical copies of a Bell
diagonal state of $2$ qubits and end up, after local operations and
classical communication, with $m<n$ Bell diagonals (possibly
statistically dependent) with higher joint fidelity than $m$ copies of
the original Bell diagonal state. For non-Bell-diagonal initial states
one can first perform $n$ separate optimal single-copy distillation
protocols to make them Bell-diagonal \cite{VDD:01}. Our protocol can
be used in a recurrence scheme followed by the hashing protocol as in
\cite{BDS:96,BBP:96}. We propose a protocol with $n=4$ and $m=1$ that
does significantly better than existing protocols. Our results can be
used to find even better protocols for other values of $n$ and $m$
that are tuned for specific initial states.

We see three main reasons for studying entanglement distillation protocols.
The first and most obvious reason is that entanglement distillation
protocols are a means of obtaining states that are closer to maximally
entangled pure states, as needed in typical applications like
teleportation, from mixed states that can be reached by sending
one qubit of an entangled pair through a realistic channel. 
A second reason to study distillation protocols is that asymptotic
protocols yield a lower bound for ``entanglement of distillation'', an
important measure of entanglement, that is in itself a lower bound for
any sensible measure of entanglement \cite{Hor:01}. In this context we
also mention the upper-bounds on entanglement of distillation obtained
in \cite{Rai:01}.
A third reason is that multi-copy entanglement
protocols can be considered as applications of entanglement, where
more can be done in the presence of entangled pairs than without. We
hope that studying these mechanisms will reveal some information on
the important problem of how exactly the presence of entanglement
enables one to do things that are impossible without.

Multi-copy mixed state entanglement distillation for qubit pairs was
first studied in \cite{BDS:96,BBP:96}. An improved variant of the
two-copy protocol in that paper was described in \cite{DEJ:96} under
the title of quantum privacy amplification.  These protocols as well as
ours start from $n$ identical qubit pairs in a Bell-diagonal state,
shared by Alice and Bob. A crucial ingredient of these protocols is a
local unitary operation, performed by Bob and Alice on their $n$
qubits, that results globally in a permutation of the $4^n$ possible
tensor products of Bell states.  The key ingredient of this paper is a
characterization, by means of a binary matrix group, of all possible
local permutations of the products of Bell states. This enables a
search for the best protocol within this setting.  In
section~\ref{seclocper} we study local permutations of products of
Bell states. In section~\ref{secprot} we discuss the new protocols. In
section~\ref{secsim} we discuss the combination of our protocols with
a recurrence scheme and the hashing protocol and show the strength of
our protocols by computer simulations.

\section{Local Permutations of products of Bell states}
\label{seclocper}
In this section we study the class of local unitary operations
that can be performed by Alice and Bob locally and result in a
permutation of the $4^n$ (tensor) products of Bell states, where $n$
is the number of qubit pairs. These local permutations are the key
ingredient of the new distillation protocols described in the next
section.

We will code products of Bell states by binary vectors by
assigning two-bit vectors to the Bell states as follows
\begin{equation}\label{binrep}
\begin{array}{lclcl}
 |\Phi^+\rangle &=& \frac{1}{\sqrt{2}}(|00\rangle+|11\rangle) &=& |B_{00}\rangle\\
 |\Psi^+\rangle &=& \frac{1}{\sqrt{2}}(|01\rangle+|10\rangle) &=& |B_{01}\rangle\\
 |\Phi^-\rangle &=& \frac{1}{\sqrt{2}}(|00\rangle-|11\rangle) &=& |B_{10}\rangle\\
 |\Psi^-\rangle &=& \frac{1}{\sqrt{2}}(|01\rangle-|10\rangle) &=& |B_{11}\rangle.
\end{array}
\end{equation}
A product of $n$ Bell states is then described by a $2n$-bit vector,
e.g. $|B_{001101}\rangle=|B_{00}\rangle|B_{11}\rangle|B_{01}\rangle=|\Phi^+\rangle
  |\Psi^-\rangle |\Psi^+\rangle$.  

We will also exploit a correspondence between Bell states and Pauli matrices
\[
\begin{array}{lcl}
 |\Phi^+\rangle &\rightarrow&\frac{1}{\sqrt{2}}
 \sigma_{00}=\frac{1}{\sqrt{2}}\sigma_0=\frac{1}{\sqrt{2}}\left[ \begin{array}{rr} 1 &
 0\\0 & 1\end{array} \right]\\ 
 |\Psi^+\rangle &\rightarrow&\frac{1}{\sqrt{2}}
 \sigma_{01}=\frac{1}{\sqrt{2}}\sigma_x=\frac{1}{\sqrt{2}}\left[ \begin{array}{rr} 0 &
 1\\1 & 0\end{array} \right]\\ 
 |\Phi^-\rangle &\rightarrow&\frac{1}{\sqrt{2}}
 \sigma_{10}=\frac{1}{\sqrt{2}}\sigma_z=\frac{1}{\sqrt{2}}\left[ \begin{array}{rr} 1 &
 0\\0 & -1\end{array} \right]\\ 
 |\Psi^-\rangle &\rightarrow&\frac{1}{\sqrt{2}}
 \sigma_{11}=\frac{1}{\sqrt{2}}\sigma_y=\frac{1}{\sqrt{2}}\left[ \begin{array}{rr} 0 &
 -i\\i & 0\end{array} \right]\\ 
\end{array}
\]
A tensor product of $n$ Bell states is then described by a tensor product of
Pauli matrices, e.g. $|\Phi^+\rangle |\Psi^-\rangle |\Psi^+\rangle \rightarrow
\frac{1}{\sqrt{8}}\sigma_0\otimes\sigma_y\otimes\sigma_x=
\frac{1}{\sqrt{8}}\sigma_{00}\otimes\sigma_{11}\otimes\sigma_{01}$.
We will also use longer vector subscripts to denote such tensor
products, e.g.
\begin{equation}\label{sigmav}
 \sigma_{001101}=\sigma_{00}\otimes\sigma_{11}\otimes\sigma_{01}. 
\end{equation}
In this representation of pure states of $2n$ qubits as $2^n\times
2^n$-matrices $\tilde\Psi$, local unitary operations $|\psi\rangle\rightarrow
(U_A\otimes U_B)|\psi\rangle$, in which Alice acts on her $n$ qubits
(jointly) with an operation $U_A$ and Bob on his with an operation
$U_B$, are represented by
\begin{equation}\label{locun}
 \tilde\Psi \rightarrow U_A \tilde\Psi U_B^T.
\end{equation}

We are now in a position to state the main result of this section:
\begin{theorem}
(i) If a local unitary operation (\ref{locun}) results in a permutation
of the $4^n$ tensor products of $n$ Bell states, this permutation can be
represented on the binary vector representations~(\ref{binrep}) as an
affine operation
\begin{equation}
\begin{array}{ll}
 & \phi: \mathbb{Z}_2^{2n}\rightarrow \mathbb{Z}_2^{2n}:x\rightarrow Ax+b\\
 \text{with } &A\in\mathbb{Z}_2^{2n\times 2n}, b\in\mathbb{Z}_2^{2n}\\
 \text{and } & A^T P A = P,\\
 \text{where } &P=\text{diag}[\left[ \begin{array}{cc}0&1\\1&0\end{array} \right],\ldots,\left[ \begin{array}{cc}0&1\\1&0\end{array} \right]]
\end{array}
\end{equation}
(ii) Conversely, any such permutation can be realized by local unitary
operations.
\end{theorem}

Note that all multiplication and addition should be done modulo 2.
We call a matrix $A$ satisfying $A^TPA=P$ $P$-orthogonal. The affine
and linear transformations considered are invertible and therefore
amount to a permutation of $\mathbb{Z}_2^{2n}$. In the sequel we sometimes
directly refer to the linear transformations as permutations.

{\bf Proof:}
We first prove part (i).
One can easily check that $\sigma_v \cdot \sigma_w \sim \sigma_{v+w}$,
where $v$ and $w$ are binary vector indices as in (\ref{sigmav}), and
the $\sim$-sign means equal up to a complex phase (in this case
$1$,$i$,$-1$ or $-i$). Such a phase is irrelevant as the Pauli matrices here
are matrix representations of pure state vectors.

Assume now that $U_A$ and $U_B$ indeed result in a permutation
$\pi:\mathbb{Z}_2^{2n}\rightarrow\mathbb{Z}_2^{2n}$, then the null vector is mapped to some
vector $v=\pi(0)$. Accordingly $U_A \sigma_0 U_B^T\sim\sigma_v$. Since
$\sigma_0$ is the identity matrix, we have $U_B\sim\sigma_v U_A^*$ where
$^*$ denotes complex conjugation. If we want to represent $\pi$ by
$x\rightarrow Ax+b$ we clearly have to choose $b=v$. Note that (\ref{locun})
now reads $\tilde\Psi \rightarrow U_A \tilde\Psi U_A^\dag \sigma_b.$

We now have to show that the permutation $\pi':x \rightarrow \pi(x)+b$, which maps
$\tilde\Psi$ to $U_A\tilde\Psi U_A^\dag$, is a linear $P$-orthogonal map
$\pi':x\rightarrow Ax$. Linearity of binary maps means that sums are mapped to
sums $\pi'(v+w)=\pi'(v)+\pi'(w)$. This is clearly true since $U_A
\sigma_{v+w} U_A^\dag\sim U_A \sigma_v \sigma_w U_A^\dag = U_A \sigma_v U
_A^\dag U_A \sigma_w U_A^\dag$. Furthermore, it can be verified using
the commutation and anticommutation laws for Pauli matrices, that
$\sigma_v$ and $\sigma_w$ are commutable operators if and only if
$v^TPw=0$. Since $\sigma_v$ and $\sigma_w$ are commutable if and only
if $U_A\sigma_v U_A^\dag$ and $U_A\sigma_w U_A^\dag$ are commutable,
it must hold that $v^T A^T P A w=v^T P w$ for all $v$ and $w$, which
proves $A^TPA=P$.

To prove part (ii), we will first consider $n=2$ and show that all
permutations $\pi:x\rightarrow Ax$ with $A^TPA=P$ can be generated with the
operations $\phi_u:\tilde\Psi\rightarrow U_u\tilde\Psi U_u^\dag$ with
$U_u=e^{i\frac{\pi}{4}\sigma_u}=\frac{1}{\sqrt{2}}(I+i\sigma_u)$ (with
$u\in\mathbb{Z}_2^4$). (This is also true for $n>2$, but generators affecting
more than $2$ qubits at a time will not be needed). Using $\sigma_v
\sigma_w=(-1)^{v^TPw}\sigma_w\sigma_v$, it can be shown that $\phi_u$
translated into the binary language results in a permutation
$\pi_u:x\rightarrow x+u(u^TPx)=(I+uu^TP)x$.

We will now show that the group of permutations generated by the
permutations $\pi_u$ is isomorphic to $S_6$, the group of all
permutations of $6$ elements. Next we will show that the group of
$P$-orthogonal $4\times 4$ matrices contains $6!=720$ elements, which
proves that all $P$-orthogonal permutations are generated.  Since no
permutation (except the identity) is commutable with all permutations,
$S_6$ is isomorphic to the group of transformations $\chi_q:S_6\rightarrow
S_6:p\rightarrow qpq^{-1}$ where $p$ and $q$ are permutations of $6$
elements. Such a transformation $\chi_q$ is completely determined by
specifying the images of the 15 commutations $p_{i,j}$, permutations
on $\{1,2,3,4,5,6\}$ that permute $i$ and $j$. This holds because any
permutation is a composition of such commutations and $\chi_q(p_1
p_2)=\chi_q(p_1) \chi_q(p_2)$. Note that the image under $\chi_q$ of a
commutation is again a commutation. As a result $S_6$ is isomorphic to
the group of permutations of $15$ elements obtained by restricting
$\chi_q$ to the commutations. We will show that this is exactly the
group of permutations generated by the generators $\pi_u$ (which can
be considered as permutations of $15$ elements as $0000$ can be left
out, being always mapped to itself). To this end we establish the
following correspondence between nonzero $4$-bit vectors and
commutations $\gamma: u\rightarrow p_{i,j}$:
\[
 \begin{array}{llll}
   & 0001\rightarrow p_{5,6}, & 0010\rightarrow p_{4,6}, &0011\rightarrow p_{4,5}, \\
0100\rightarrow p_{2,3}, &0101\rightarrow p_{1,4}, & 0110\rightarrow p_{1,5}, &0111\rightarrow
p_{1,6}, \\ 
1000\rightarrow p_{1,3}, &1001\rightarrow p_{2,4}, & 1010\rightarrow p_{2,5}, &1011\rightarrow
p_{2,6}, \\ 
1100\rightarrow p_{1,2}, &1101\rightarrow p_{3,4}, & 1110\rightarrow p_{3,5}, &1111\rightarrow
p_{3,6}. \\
 \end{array}
\]
It can be verified that $\gamma(\pi_u(x))=\chi_{\gamma(u)}(\gamma(x))$
for all $u$ and $x$. So $\pi_u$ and $\chi_{\gamma(u)}$ realize the
same permutation of $15$ elements. As a consequence, also products 
$\pi_{u_1}\cdot\ldots\cdot\pi_{u_k}$ realize the same permutations as
products $\chi_{\gamma(u_1)}\cdot\ldots\cdot\chi_{\gamma(u_k)}$. This
finally establishes the isomorphism between $S_6$ and the permutations
generated by the $\pi_u$.

It remains to be shown that there are $6!$ $P$-orthogonal $4\times 4$
matrices. It follows from $A^TPA=P$ that $A$ is $P$-orthogonal if and
only if all pairs of columns of $A$ represent commutable
$\sigma_{a_i}$ except for the first and second or the third and fourth
column. Therefore to make an arbitrary $P$-orthogonal matrix, the
first column $a_1$ can be chosen to be any nonzero 4-bit vector ($15$
choices), the second column should satisfy $a_1^TPa_2=1$ (one linear
condition yielding $8$ possible $a_2$), the third column should be
commutable with $a_1$ and $a_2$ (two linear conditions yielding $3$
choices after excluding $0000$) and finally the fourth column should
be commutable with $a_1$ and $a_2$ and noncommutable with $a_3$ (three
linear conditions, yielding $2$ possibilities). This results in
$15 \cdot 8 \cdot 3 \cdot 2=720=6!$ possibilities. This ends the proof
for $n=2$.

For $n>2$ we turn to the matrix picture and show that every
$P$-orthogonal matrix $A$ can be reduced to the identity matrix by
two-qubit operations, i.e. $4\times 4$ $P$-orthogonal matrices
embedded in an identity matrix on rows and columns $2k+1, 2k+2, 2l+1,
2l+2$ for some $k,l\in\{0,\ldots,n-1\}$. We concentrate on two columns
of $A$ at a time, first $1$ and $2$, then $3$ and $4$ and so on and
transform them to the corresponding columns of the identity matrix
with two-qubit operations. Assume, without loss of generality, that we
are working on columns $1$ and $2$, then we name
$K^{(k,l)}=A_{\{2k+1,2k+2,2l+1,2l+2\},\{1,2\}}$. If the two columns of
$K^{(k,l)}$ are commutable they can be thought of as the first and third
column of a $4\times 4$ $P$-orthogonal matrix and can be reduced by a
two-qubit operation to the first and third column of an identity
matrix. If the two columns of $K^{(k,l)}$ are noncommutable they can be
reduced to the first and second column of an identity matrix. One can
see that by combining such two qubit operations the first two columns
of $A$ can be reduced to the first two columns of an identity
matrix. Due to the commutability relations between the columns of $A$,
as a result, also the first two rows become the first rows of an
identity matrix. One can now proceed in a similar way with the next
pairs of columns until the whole matrix is reduced to the identity
matrix. The composition of the inverses of all two-qubit operations
that were applied yields a decomposition of $A$ into two-qubit
operations that can be realized by local unitary operations as shown
above. This ends the proof. \hfill $\square$

In the proof we saw that linear transformations ($b=0$) correspond to
operations with $U_B=U_A^*$, i.e. $\tilde\Psi\rightarrow U_A\tilde\Psi
U_A^\dag$. The matrices $U_A$ that under this action map tensor
products of Pauli matrices to tensor products of Pauli matrices
possibly with a minus-sign are known to form the Clifford group,
studied in \cite{Got:98,Got:97} in the context of quantum error
correction and quantum computation. The P-orthogonal matrices form a
group that is isomorphic to a quotient group of the Clifford group.
The Clifford group is known to be generated by CNOT operations and
one-qubit operations that map Pauli matrices to Pauli matrices. It is
possible that this knowledge may be used to give other proofs for the
theorem above. However, we think that our set of generators and
the isomorphism between $P$-orthogonal $4\times 4$ matrices and
permutations of $6$ elements are worthwhile results in their own. It
also follows that the CNOT operation should be decomposable in terms
of our generators (at least up to phase factors, but we can do
better). One can easily verify that
$\text{CNOT}=\frac{1+i}{\sqrt{2}}e^{-i\frac{\pi}{4}\sigma_{1000}}
e^{i\frac{\pi}{4}\sigma_{1001}}e^{-i\frac{\pi}{4}\sigma_{0001}}$. Note
that the first and last operation are actually 1-qubit operations.

\section{Mixed state multi-copy entanglement distillation\protect\\ from pairs
  of qubits}
\label{secprot}

The distillation protocols presented in this paper can be summarized
as follows. 
\begin{enumerate}
\item Start from $n$ identical independent Bell diagonal states with
  entanglement. This yields a mixture of $4^n$ tensor products of Bell states.
\item Apply a local permutation of these $4^n$ products of Bell states
  as described in the previous section. As a result the $n$ qubit pairs get 
  statistically dependent.
\item Check whether the last $n-m$ qubit pairs are
  $|\Phi\rangle$-states ($|\Phi^+\rangle$ or $|\Phi^-\rangle$). This can be
  accomplished locally by measuring both qubits of each pair in the
  $|0\rangle,|1\rangle$ basis, and checking whether both measurements yield
  the same result. 
\item If all measured pairs were $|\Phi\rangle$-states, keep the first $m$
  pairs. This is a new mixture of $4^m$ products of Bell states.
\end{enumerate}

This is a generalization of a protocol with $n=2$ and $m=1$,
presented in \cite{BDS:96,BBP:96}. In that protocol the applied local
permutation consisted of a bilateral CNOT operation by Alice and Bob.
In our protocol, we will only consider linear permutations ($b=0$) as
one can easily see nothing can be gained by considering affine
permutations. In the next section we discuss how to choose the local
permutation so as to obtain a good protocol. The main result of this
section is a formula for the resulting state of $m$ pairs as a
function of the permutation of Bell states performed in step 2 of the
protocol:

\begin{theorem}
\label{thprot}
If Alice and Bob apply the above protocol, starting from $n$
independent identical copies of a Bell diagonal state
$p_{00}|\Phi^+\rangle\langle\Phi^+|+ p_{01}|\Psi^+\rangle\langle\Psi^+|+
p_{10}|\Phi^-\rangle\langle\Phi^-|+ p_{11}|\Psi^-\rangle\langle\Psi^-|$ with
$p_{00}\geq p_{01}\geq p_{10} \geq p_{11}$, and with entanglement,
i.e. $p_{00}>\frac{1}{2}$, with in step $2$ a local permutation,
$\pi:x\rightarrow Ax$ with $A^TPA=P$, the resulting state of the remaining
$m$ qubit pairs is given by
\begin{equation}
\label{eqes}
 2^{n-m}\sum_{y\in\mathbb{Z}_2^{2m}} \left(
   \frac{\sum_{x\in{\cal S}+PA^TP\bar y} p_{x}}
        {\sum_{x\in{\cal S}} s_{x}}
 \right) |B_y\rangle\langle B_y|
\end{equation}
where ${\cal S}$ is the subspace spanned by the rows of $AP$ with
indices $2m+2, 2m+4,\ldots,2n$,
\[
 \left[ \begin{array}{c}s_{00}\\s_{01}\\s_{10}\\s_{11}\end{array} \right]=
 \left[ \begin{array}{rrrr}1&1&1&1\\ 1&1&-1&-1\\ 1&-1&1&-1\\ 1&-1&-1&1\end{array} \right]
 \left[ \begin{array}{c}p_{00}\\p_{01}\\p_{10}\\p_{11}\end{array} \right],
\]
$\bar y$ is $y$ extended with $2(n-m)$ zeros, and the long vector
indices of $p$ and $s$ and $B$ behave like the indices of $\sigma$ in
the previous section, e.g. $p_{001101}=p_{00}p_{11}p_{01}$.
\end{theorem}

{\bf Proof:}
After the permutation, and before the measurement, the state of the
$n$ qubit pairs is given by $\sum_{x\in\mathbb{Z}_2^{2n}} p_x |B_{Ax}\rangle\langle
B_{Ax}|$. The states $|B_{Ax}\rangle$ with
$(Ax)_{2m+2},(Ax)_{2m+4},\ldots,(Ax)_{2n}=0$ yield $|\Phi\rangle$-states and
will be kept. These are the states $|B_{Ax}\rangle$ for which $x$ is commutable
with the rows $a_{2m+2},\ldots,a_{2n}$ of $AP$. If we call the subspace
of these vectors $x$, ${\cal R}$, the success rate (probability of keeping
the first $m$ pairs) is $\sum_{x\in{\cal R}}p_x$. Among the states
that are kept, the ones with $(Ax)_j=y_j, j=1,\ldots,2m$ yield
$|B_y\rangle$-states. Together with the conditions for
being kept, these are $2m+(n-m)$ independent linear conditions,
yielding a coset of an $(n-m)$-dimensional subspace of
$\mathbb{Z}_2^{2n}$. This subspace must be ${\cal S}$ since the latter is
$(n-m)$-dimensional and satisfies all homogeneous conditions (with
$y=0$) by the $P$-orthogonality of $A$. The right coset is obtained
by adding $PA^TP\bar y$ (a combination of the first $2m$ rows of $AP$,
determined by $y$). As a result the state of the first $m$ pairs after
the measurement is

\[
\sum_{y\in\mathbb{Z}_2^{2m}} \left(
   \frac{\sum_{x\in{\cal S}+PA^TP\bar y} p_x}
        {\sum_{x\in{\cal R}} p_x}
  \right) |B_y\rangle\langle B_y|
\]

If all coefficients (for all $y$) are calculated, the denominator
$\sum_{x\in{\cal R}} p_x$ can be calculated as the sum of the $2^{2m}$
numerators. If only one coefficient is needed (for instance if only
the fidelity of the end state is needed) the denominator can be
calculated in a more efficient way as $\sum_{x\in{\cal R}}
p_x=2^{-(n-m)}\sum_{x\in{\cal S}} s_x$. One can easily verify that
$s_v= \sum_{x\in\mathbb{Z}_2^{2n}}(-1)^{v^TPx} p_x$ (First verify for 2 bits
and then extend). Therefore $\sum_{v\in {\cal S}}s_v=
\sum_{x\in\mathbb{Z}_2^{2n}}(\sum_{v\in {\cal S}}(-1)^{v^TPx}) p_x$. If $x$
commutes with all $v\in {\cal S}$, $(-1)^{v^TPx}=1$ for all $v\in
{\cal S}$ and $\sum_{v\in {\cal S}}(-1)^{v^TPx}=2^{n-m}$. If $v\not\in
{\cal S}$, one can easily show that half the coefficients
$(-1)^{v^TPx}$ are $1$ and half are $-1$.  Now the states $x\in{\cal
  R}$ are exactly the ones for which $x$ is commutable with all
elements of ${\cal S}$. Therefore $\sum_{x\in{\cal S}} s_x=
2^{n-m}\sum_{x\in{\cal R}}p_x$. This concludes the proof.
\hfill$\square$

\section{Recurrence schemes}
\label{secsim}

With the above formula for the end state of the protocol
(Theorem~\ref{thprot}), it is 
possible to derive good protocols by searching over all possible
values for the relevant rows of the $P$-orthogonal matrix $A$ and
optimizing some quality measure. Typically this measure will depend
on the fidelity of the end state and the success rate of the protocol
(the probability of having $|\Phi\rangle$-states in the measured pairs).
In that case, one only needs the first coefficient (the fidelity) and
the denominator (the success rate) in (\ref{eqes}), which both only
depend on ${\cal S}$, a space spanned by only $n-m$ rows of ${\cal S}$.
Although this drastically limits the search space, it still grows
exponentially with growing $n$.

Therefore, to come up with schemes for large $n$, one needs to use the
recurrence scheme, as was proposed for $n=2$ and $m=1$ in
\cite{BDS:96,BBP:96}. If $m=1$, this scheme means that the above
protocol is performed $n$ times (with the same local permutation) and
the $n$ identical end states are taken as the input for a new step. Of
course more than two steps are possible too. One could also envision
recurrence schemes with $m\neq 1$, for instance, combining two
end states of an $n=4$,$m=2$-protocol to yield the input for a second
step with $n=4$. In that case however the input for the second step
would no longer consist of $n$ independent pairs. Although this only
requires a minor modification of the above results ($p_x$ and $s_x$ can
no longer be interpreted as products of $p_{00},\ldots,p_{11}$), we
will not consider this case in this paper. 

To end up with almost pure Bell states, the recurrence scheme can best
be combined with the hashing protocol as in \cite{BDS:96,BBP:96}. The
hashing protocol is the best known asymptotic protocol (for
$n\rightarrow\infty$) but can only be applied to Bell diagonal states with
high enough fidelity. The combined protocol then consists of first
applying a few recurrence steps and then switching to the hashing
protocol. 

The best known $n=2$,$m=1$-recurrence scheme is the one of
\cite{DEJ:96}. In our language it amounts to a scheme with a
$4\times 4$ $P$-orthogonal matrix whose last line is $11\,11$. It can
be proven that this scheme yields the best achievable fidelity after
one step (though not achieved with the best success rate) for initial
probabilities that are ordered $p_{00}>p_{01}\geq p_{10}\geq
p_{11}$. For this reason it is also best to apply a pair per pair
transformation after each recurrence step, which reorders the
probabilities of the end state if they are not ordered. (One can
easily find such one-pair transformations using the theory of
section~\ref{seclocper} or equivalently using the local operations of
\cite{BDS:96,BBP:96}.) 

Although it is probably best to search for a new protocol for every given
initial state, we propose below a protocol which we think is good
if one does not have the time for such a search. We show by computer
simulations that it performs better than the $n=2$ scheme. 

Our scheme is an $n=4$,$m=1$ recurrence scheme combined with hashing,
and with as the last step possibly an $n=2$,$m=1$-step if this can
lead to better performance. For the local permutation (determined by
the $P$-orthogonal matrix $A$) we choose a permutation that is found
experimentally to often lead to the best fidelity after one step, when
starting with ordered probabilities. For this reason we also apply a
reordering in between recurrence steps as discussed for the
$n=2$-protocol above. The chosen local permutation corresponds to an
$8\times 8$ $P$-orthogonal matrix $A$ whose fourth, sixth and eighth
row span the space spanned by $\{10\,11\,11\,10, 01\,10\,11\,00,
11\,10\,10\,11\}$. This can be achieved by the operations
\begin{equation}
\label{equa}
\begin{array}{ll}
 U_A=U_B^*= & e^{i\pi/4\sigma_{10\,01\,00\,00}}
              e^{i\pi/4\sigma_{01\,00\,00\,01}}\\
            & e^{i\pi/4\sigma_{10\,00\,11\,00}}
              e^{i\pi/4\sigma_{00\,01\,10\,00}}.
\end{array}
\end{equation}
In this realization the first and second row of the $P$-orthogonal
matrix $A$ are $01\,10\,00\,10$ and $10\,10\,10\,10$.
These rows are needed to compute the reordering operations between the
steps, for although the three values of $p'_{01}$, $p'_{10}$ and $p'_{11}$
after one step of the protocol are fixed, their order is not. (The
three cosets of ${\cal S}$ in ${\cal R}$ in equation~(\ref{eqes}) are 
fixed but not their order).

This realization was found by
exhaustive search over all operations that can be realized by $4$
consecutive elementary two-qubit operations. If, for protocols with
larger $n$ for instance, no such simple realization can be found in a
reasonable amount of time, one can always find a realization using the
theory of section~\ref{seclocper} but this can increase
the total amount of work for the distillation protocol. This was also
one of the reasons for choosing $n=4$ in the proposed protocol.

\begin{figure}
\begin{center}
 \epsfig{file=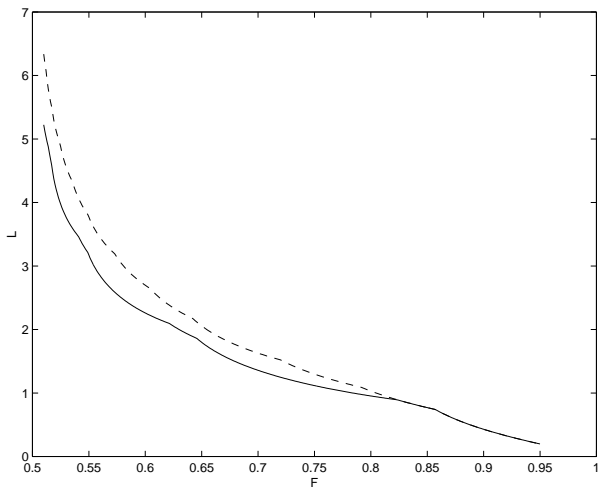,width=8cm}
\end{center}
\caption{Comparison of 10-logarithm of inverse asymptotic yield L for
  input Werner states with fidelity F for proposed protocol (full
  line) and existing recurrence/hashing protocol (dashed line)}
\label{figcomp}
\end{figure}

As a performance measure we have chosen the expected average number of
input pairs needed per output Bell state in an asymptotic protocol
(the inverse of the asymptotic yield). The number of recurrence steps
was also chosen as to optimize this measure. Fig.~\ref{figcomp} shows
the performance for our method ($n=4$,$m=1$-recurrence with the local
permutation realized by $U_A$ as in (\ref{equa}), with reordering
between the steps, possibly one last $n=2$,$m=1$-step, and optimal
switching to hashing protocol) and the method of \cite{DEJ:96} with
reordering between the steps and optimal switching to the Hashing
protocol. For the sake of simplicity the figure only shows the results
for Werner states (with $p_{00}=F>\frac{1}{2}$ and
$p_{01}=p_{10}=p_{11}=\frac{1-F}{3}$), but the method also performs
better for non-Werner states.

To do better than this protocol for a specific initial Bell diagonal
state, one can do several things depending on the amount of computing
time available. One can try recurrence schemes with higher $n$ and
even higher $m$, but the amount of time needed increases fast with
increasing $n$. There is of course no obligation to take the same
local permutation in consecutive recurrence steps. One can also
consider distilling more than one end state at once. Making two states
with two $n=4$,$m=1$-protocols is just a special case of a non-optimal
$n=8,m=2$-protocol. One can of course search for better ones if one
has the time. In this case, the two obtained Bell states will not be
independent but as the fidelity goes to $1$, their dependence will vanish.
Also two consecutive recurrence steps, say two $n=2$,$m=1$-steps, can
be considered as one bigger non-optimal step, in this case with
$n=4$,$m=1$. So if one has the time one can in theory always go for a
one shot protocol (no recurrence), but if one combines with the
recurrence scheme one can always afford lower initial entanglement
with the same amount of computing time.

\section{Conclusion}
We have derived new protocols for distillation of entanglement from
mixed states of two qubits. The protocols were based on a
characterization of the group of all locally realizable permutations
of the $4^n$ possible tensor products of $n$ Bell states. Our
protocols perform significantly better than existing protocols as was
shown by computer simulation. We also indicated how to derive even
better protocols for specific initial states.

\begin{acknowledgments}
Our research is supported by grants from several funding agencies and sources:
Research Council KULeuven: Concerted Research Action GOA-Mefisto 666
(Mathematical Engineering), several PhD/postdoc \& fellow grants; 
Flemish Government: Fund for Scientific Research Flanders (several
PhD/postdoc grants, projects G.0256.97 (subspace), G.0240.99
(multilinear algebra), research communities ICCoS, ANMMM)); 
Belgian Federal Government: DWTC (IUAP IV-02 (1996-2001) and IUAP V-22
(2002-2006): Dynamical Systems and Control: Computation,
Identification \& Modelling); The European Commission: Esprit project:
DICTAM
\end{acknowledgments}

\bibliography{dist}

\end{document}